\documentclass[]{pasj00}
\usepackage{natbib}
\draft

\begin{document}
\SetRunningHead{H. Bak{\i}\c{s} et al.}{Ori~OB1a Member: IM~Mon}
\Received{}%{yyyy/mm/dd}
\Accepted{}%{yyyy/mm/dd}
\Published{}%{yyyy/mm/dd}

\title{Study of Eclipsing Binary and Multiple Systems in OB Associations: I. Ori~OB1a - IM~Mon}

\author{%
Hicran \textsc{BAKI\c{S}}\altaffilmark{1}
Volkan \textsc{BAKI\c{S}}\altaffilmark{1}
Sel\c{c}uk \textsc{B\.{I}L\.{I}R}\altaffilmark{2}
Zden\v ek \textsc{MIKUL\'A\v SEK}\altaffilmark{3,4}
Miloslav \textsc{ZEJDA}\altaffilmark{3}
Esma \textsc{YAZ}\altaffilmark{2}
Osman \textsc{DEM\.{I}RCAN}\altaffilmark{1}
and
\.{I}brahim \textsc{BULUT}\altaffilmark{1}}
\altaffiltext{1}{\c{C}anakkale Onsekiz Mart University, Physics Department and Ulup{\i}nar Observatory, Terzio\v{g}lu Campus, TR-17020, \c{C}anakkale, Turkey}
\email{bhicran@comu.edu.tr}
\email{bakisv@comu.edu.tr}
\email{demircan@comu.edu.tr}
\email{ibulut@comu.edu.tr}
\altaffiltext{2}{\.{I}stanbul University, Science Faculty, Department of Astronomy and Space Sciences, 34119 University-\.{I}stanbul, Turkey}
\email{sbilir@istanbul.edu.tr}
\email{esmayaz@istanbul.edu.tr}
\altaffiltext{3}{Masaryk University, Department of Theoretical Physics and Astrophysics, Brno, Czech Republic}
\altaffiltext{4}{ V\v SB -- Technical University, Observatory and Planetarium of J. Palisa, Ostrava, Czech Republic}
\email{mikulas@physics.muni.cz}
\email{zejda@physics.muni.cz}


\maketitle

\begin{abstract}
All available photometric and spectroscopic observations were collected and used as the basis of a
detailed analysis of the close binary IM~Mon. The orbital period of the binary was refined to
1.\timeform{d}19024249(0.00000014). The Roche equipotentials, fractional luminosities (in
\textit{B, V} and \textit{H$_p$} - bands) and fractional radii for the component stars in addition
to mass ratio $q$, inclination $i$ of the orbit and the effective temperature $T_{\rm eff}$ of the
secondary cooler less massive component were obtained by the analysis of light curves. IM~Mon is
classified to be a detached binary system in contrast to the contact configuration estimations in
the literature. The absolute parameters of IM~Mon were derived by the simultaneous solutions of
light and radial velocity curves as M$_{\rm 1,2}=5.50(0.24)$ M$_{\odot}$ and 3.32(0.16)
M$_{\odot}$, R$_{\rm 1,2}=3.15(0.04)$ R$_{\odot}$ and 2.36(0.03) R$_{\odot}$, $T_{\rm
eff1,2}=17500(350)$~K and 14500(550)~K implying spectral types of B4 and B6.5 ZAMS stars for the
primary and secondary components respectively. The modelling of the high resolution
spectrum revealed the rotational velocities of the component stars as V$_{\rm
rot1}=147(15)$ km\,s$^{-1}$ and V$_{\rm rot2}= 90(25)$ km\,s$^{-1}$. The photometric distance of
353(59) pc was found more precise and reliable than Hipparcos distance of 341(85) pc. An
evolutionary age of 11.5(1.5) Myr was obtained for IM~Mon. Kinematical and dynamical analysis
support the membership of the young thin-disk population system IM~Mon to the Ori~OB1a association
dynamically. Finally, we derived the distance, age and metallicity information of Ori~OB1a
sub-group using the information of IM~Mon parameters.
\end{abstract}

\section{Introduction}

Young stellar associations ($\leq$~50 Myr) and open clusters have a crucial importance in advancing
our understanding of star formation and the first stages of stellar evolution.  Since the
galactic acceleration does not have a chance to affect the kinematical properties of these
young stellar groups, the stellar content of an association is preserved. Consequently one
can obtain kinematical, dynamical and chemical properties of these young stellar groups by studying
their secure members. Today nearby associations within the Solar neighbourhood have been
very well identified. Their stellar content of each association has been
precisely determined up to a magnitude limit of $V\sim$10.5 mag using astrometric data of the
Hipparcos satellite \citep[i.e.][]{dezeeuw1999, hoogerwerf2000, melnik2009}. In addition
the new reduced Hipparcos catalogue \citep{vanleeuwen} gave an opportunity to investigate
the astrometric data of the stellar content of a large number of open clusters, associations and
moving groups more accurately. The first application of the new reduced Hipparcos astrometric data
is applied to stellar groups within the solar neighbourhood by \citet{melnik2009}. This
and forthcoming studies of young stellar groups will improve our understanding of the
history of star formation, the initial mass function, chemical and dynamical evolution of the
Milky Way.

Recent statistical studies, such as \citet{brown2001}, \citet{bouy2006} and
\citet{kouwenhoven2007} show a high ratio of multiplicity in stellar formation regions (SFRs) and
claim that it is not a coincidence, but a characteristic of star formation. The detailed
study of multiple systems (especially those with eclipsing components) in SFRs
will reveal the fundamental stellar parameters more directly and with higher precision compared to
those obtained from single stars and thus impose more stringent tests on stellar evolution.
Critical tests of stellar evolution require masses and radii with a precision better than 3 per
cent (i.e. \citet{andersen, torres2010}). However, methods developed for single stars are not
capable of delivering masses with a precision better than 5 per cent and radii remain uncertain by
a factor of 1.5. Consequently studying single stars does not enable us to obtain accurate
dimensions and, therefore, to test the most recent evolutionary models. Recent studies on $\eta$
Mus by \citet{bakis2007} in Lower Centaurus-Crux association, on V578 Mon by
\citet{pavlovski2000,pavlovski2005} in NGC 2244, and on AB Dor by \citet{luhman2006} in the AB Dor
association, demonstrate the precision with which age, chemical composition and
kinematical properties can be determined by studying such high-mass systems.

In the present study, we analyzed the high resolution spectra and \textit{BVH$_p$}
photometric data of IM~Mon, which is located in the region of Ori~OB1a association. IM~Mon
is a bright ({\rm V}$\sim$6.5 mag), early-type (({\it B-V})=--0.14 mag) and short orbital
period ({\rm P$\sim$1.2 days}) eclipsing binary system. Its spectroscopic and photometric
variations were discovered by \citet{pearce} and \citet{gum} respectively. The eccentricity of the
spectroscopic orbit obtained by \citet{pearce} was commented to be spurious by \citet{cester} in
their study on the determination of photometric elements of 14 detached systems. \citet{cester}
studied early photometric observations of IM~Mon, which were collected by \citet{gum} in integral
light and by \citet{sanyal1965} in $B$ and $V$ filters. However, due to a large scatter in all
photometric observations, which is attributed to the intrinsic variability of one of the
components by \citet{sanyal1965}, none of the authors was able to find a unique and precise
solution for the system. A recent spectroscopic study of \citet{bakis2010} revealed the
spectroscopic orbital elements of IM~Mon and showed that its orbit is circular.

In order to reveal more precise absolute dimensions of IM~Mon and to test its membership to
Ori~OB1a, we included it into our list of eclipsing binaries in the region of OB associations.
Using all literature based data the orbital period of IM~Mon is revised in \S2.3. High
resolution spectral lines of IM~Mon are modelled and atmosphere parameters are derived in \S3. The
close binary stellar parameters of the system are determined by the analysis of light and
radial velocity (RV) curves in \S4. In \S5 the absolute parameters of the components are
derived together with the age and distance of the system. This information enabled us to establish the
absolute dimensions of the close binary and properties of the Ori~OB1a association through the
kinematical and dynamical properties of IM~Mon. Finally we summarized our study and
present our conclusions in \S6.

\section{Observational Data}
We collected as many original individual measurements of IM Mon as possible from the literature. In table \ref{Listsources}, all available photometric and spectroscopic observations of IM~Mon are listed. All the data given in table \ref{Listsources} are used for ephemeris
determination for O--C analysis, whereas only relatively more precise photometric data are used for
light curve (LC) modelling. As a starting orbital period, we adopted
$P=1\timeform{d}.1902424$, which is published by \citet{kreiner} and later used by
\citet{bakis2010} for their radial velocity analysis.

\begin{table*}
\begin{center}
\caption{List of sources of measurements used (after removing outliers). C(B) and C(R) stand for clear filters with B and R effective bands, respectively.\label{Listsources}}
\begin{tabular}{cccccc}
\hline
 No. & Source & Filter & No. of points & $\sigma$  & Notes \\
  \hline
Photometry & & & & [mmag] & \\
  1  & Gum (1951)                & \textit{C(B)}  & 33  & 3.7 & normal points \\
  2  & Sanyal et al. (1965)      & \textit{B}     & 42  & 3.4 & normal points \\
  3  &                           & \textit{V}     & 41  & 4.4 & normal points \\
  4  & Shobbrook (2004)          & \textit{V}     & 40  & 7 & \\
\hline
  5  & Hipparcos           & \textit{Hp}    & 118 & 6 & \\
  6  &                     & $V_{\rm{T}}$   & 159 & 38 &\\
  7  &                     & $B_{\rm{T}}$   & 159 & 41 &\\
  8  & ASAS                & \textit{V}     & 435 & 26 &\\
  9  & Pi                  & \textit{C(R)}  & 490 & 32 &\\
 \hline
Spectroscopy & & & & [km\,s$^{-1}$] & \\
  10 & Pearce (1932)             & RV    & 18  & 29 & primary \\
  10 & Pearce (1932)             & RV    & 13  & 41 & secondary \\
  11 & Bak{\i}\c{s} et al. (2010)& RV    & 23  & 13 & primary \\
  11 & Bak{\i}\c{s} et al. (2010)& RV    & 20  & 16 & secondary \\
\hline \\
\end{tabular}
\end{center}
\end{table*}

\subsection{Photometric Data}
\begin{itemize}

  \item \citet{gum} -- Data were obtained with nine-inch refractor of the Commonwealth Observatory, Canberra equipped with a photoelectric photometer with 1P21 photomultiplier tube on effective wavelength 440.0 nm (without filter) in one season 1949-1950 (normal points related to JD 2\,433\,402); HD~45321, HD~44756 were used as comparison stars.

  \item \citet{sanyal1964} -- \textit{B} and \textit{V} observations of \textit{UBV} system (Johnson and Morgan 1951) were made during 29 nights of three seasons in 1960-1964 (normal points related to JD 2\,438\,384) at Uttar Pradesh State observatory, Naini Tal using 10 inch Cooke refractor and 15-inch reflector equipped with unrefrigerated 1P21 photomultiplier. BD
-2$^{\circ}$ 1601 (=HD 45139) and BD -3$^{\circ}$ 1414 (=HD 44720) were used as comparison and
check stars respectively.

  \item \citet{shobbrook} -- The 24-inch (61-cm) telescope of the Australian National University at Siding Spring Observatory with photometer -- a cooled GaAs photomultiplier and the Motorised Filter Box with Str\"{o}mgren $y$ and \textit{b} filters were used in JD 2\,450\,458--2\,452\,026. The observations were reduced to the Johnson \textit{V} scale using $uvby$ secondary standard stars of Cousins (1987) in the E Regions (sic). HR~2325 (=HD 45321) and HR~2344 (=HD 45546) were used as comparison stars.

  \item Hipparcos \citep{esa} -- Observations were made by Hipparcos satellite equipment (reflector 29-cm, f=1.4-m) in the interval JD 2\,447\,960--2\,449\,058. More details are on webpage. \footnote{http://cadcwww.dao.nrc.ca/astrocat/hipparcos/}

  \item ASAS \citep{pojm97} -- Observations were obtained from two ASAS observing stations, one is in Las Campanas Observatory, Chile (since 1997)   and the other is on Haleakala, Maui (since 2006) between JD 2\,452\,731--2\,455\,167. Both were equipped with two wide-field instruments (72-mm, f=0.2-m), observing simultaneously in \textit{V} and $I$ band. However, only \textit{V} measurements are available for IM~Mon. More details and data archive are on webpage.\footnote{http://www.astrouw.edu.pl/asas/}

  \item Pi of the Sky \citep{malek} -- The Pi of the Sky robotic telescope has been designed for monitoring of a significant fraction of the sky with good time resolution and range. Final detector consists of two sets of 16 cameras, one camera covering a field of view of 20$^\circ \times 20^\circ$. The final system is currently under construction. Required hardware and software tests have been performed with a prototype located in Las Campanas Observatory in Chile since June 2004. The set of IM~Mon measurements cover time the interval of JD 2\,453\,954--2\,454\,946. More details and data archive are on webpage.\footnote{http://grb.fuw.edu.pl/}

\end{itemize}

\subsection{Spectroscopy}

\begin{itemize}

\item \citet{pearce} -- 19 spectrograms were obtained at Dominion Astrophysical Observatory using a 72-inch telescope in the interval JD
2\,424\,942--2\,425\,682, mostly with the short-focus camera having a dispersion 49
{\AA}/mm at H$\gamma$. From these spectrograms, a total of 31 RVs (18 for primary and 13 for
secondary component) have been measured by \citet{pearce}. We used all RVs given by \citet{pearce}
in this study.

\item \citet{bakis2010} -- The spectra were taken with High Efficiency and Resolution Canterbury University
Large Echelle Spectrograph (HERCULES) of the Department of Physics and Astronomy, New Zealand. It
is a fibre-fed \'echelle spectrograph, attached to the 1-m McLellan telescope at the  Mt John
University Observatory (MJUO). All spectra were collected using the 4k $\times$ 4k
Spectral Instruments 600S CCD camera.  All 23 spectra were obtained in August 2006 (JD
2\,453\,981-2\,453\,992). A detailed description including journal of observations is
given in \citet{bakis2010}.

\citet{bakis2010} used a two-dimensional cross-correlation technique for measuring RVs of the
components of IM~Mon, whose spectra show evidently blended lines at specific orbital phases. This
technique is more suitable for the measuring RVs of close binary systems in the case of
blended spectral lines at some phases, though is not as powerful as spectra disentangling
\citep[][]{simonsturm, hadrava}. However, due to very strong blending near the eclipse
phases when the eclipsed component's spectral lines are invisible, it was not possible for them to
measure a reliable RV of the secondary component at four phases \citep[][]{bakis2010}. Notice that
in table 2 of \citet{bakis2010}, the listed RVs are uncorrected for barycentric motion
although the spectroscopic orbital elements given were obtained from the corrected RVs.
Herewith we re-list the corrected RVs in table \ref{crrvs}.

\end{itemize}

\begin{table*}
\begin{center}
\caption{RVs of IM~Mon. Orbital phases ($\varphi$) were calculated using the same ephemeris given by \citet{bakis2010}.\label{crrvs}}
\begin{tabular}{cccrccccc}
\hline
No  &  HJD        & Phase & RV${\rm _1}$ & (O-C)${\rm _1}$ & RV${\rm _2}$  & (O-C)${\rm _2}$ \\
    &  (-2453900) &($\varphi$)& (km\,s$^{-1}$)  & (km\,s$^{-1}$) & (km\,s$^{-1}$) & (km\,s$^{-1}$) \\
\hline
1   & 81.1889 & 0.866 & 133.2  & 8.8   & -153.4 & -2.7  \\
2   & 81.2155 & 0.888 & 121.4  & 11.8  & -131.4 & -5.2  \\
3   & 81.2402 & 0.909 & 118.0  & 21.8  & -96.5  & 7.6   \\
4   & 82.2019 & 0.717 & 160.2  & 4.9   & -206.5 & -4.9  \\
5   & 82.2313 & 0.742 & 165.7  & 7.3   & -197.7 & 9.0   \\
6   & 85.1955 & 0.232 & -112.7 & 3.7   & 252.4  & 5.6   \\
7   & 86.1892 & 0.067 & 5.9    & 40.9  & -      & -     \\
8   & 86.2328 & 0.104 & -43.9  & 19.6  & 139.7  & -30.4 \\
9   & 86.2442 & 0.113 & -51.1  & 18.8  & 151.4  & -18.7 \\
10  & 88.1948 & 0.752 & 163.1  & 4.5   & -203.8 & 3.2   \\
11  & 88.2354 & 0.786 & 161.0  & 5.8   & -203.4 & -1.9  \\
12  & 89.1573 & 0.561 & 53.2   & -19.3 & -      & -     \\
13  & 89.1868 & 0.585 & 91.6   & 4.1   & -112.2 & -16.4 \\
14  & 89.2141 & 0.608 & 115.5  & 7.3   & -135.4 & -11.5 \\
15  & 89.2404 & 0.631 & 123.0  & 1.2   & -162.0 & -15.6 \\
16  & 90.1456 & 0.391 & -69.5  & -2.6  & 147.8  & -17.2 \\
17  & 90.1730 & 0.414 & -52.6  & -0.9  & 111.5  & -13.4 \\
18  & 90.2098 & 0.445 & -7.4   & 19.9  & -      & -     \\
19  & 91.1578 & 0.241 & -120.0 & -2.9  & 237.9  & -10.0 \\
20  & 91.2020 & 0.279 & -103.7 & 11.6  & 253.8  & 9.0   \\
21  & 92.1558 & 0.080 & -57.4  & -11.2 & 119.2  & -11.5 \\
22  & 92.2077 & 0.123 & -76.5  & -0.4  & 202.8  & 22.6  \\
23  & 92.2390 & 0.150 & -82.1  & 9.0   & 212.7  & 7.8   \\
\hline \\
\end{tabular}
\end{center}
\end{table*}

\section{Modelling Spectral Lines}

We have modelled the spectral orders extracted from a high resolution \'echelle
spectrum of IM~Mon with the theoretical atmosphere models of \citet{kurucz1993} to obtain the
atmosphere parameters such as metallicity $[Fe/H]$, effective temperature ($T_{\rm eff}$), surface
gravity ($\log g$), projected rotational velocity (V$_{rot}$~sin$i$) and microturbulance velocity
($\zeta$) of the components. These atmosphere parameters, especially the metallicity,
which can be obtained primarily from the observed spectrum, are very useful during the
construction of the isochrones of IM~Mon. The atmosphere models we used are originally provided for
a wide range of temperature, surface gravity and metallicity by \citet{kurucz1993}. For specific
values of these parameters, one should use routines to obtain desired atmosphere parameters. We
used ATLAS9 and SYNTHE routines (Kurucz, 1993) with new opacity distribution functions (ODF)
provided by \citet{castelli2001}. In order to find the best fitting atmosphere parameters,
the Grid-Search Method (Bevington, 2003) was used. The Grid-Search Method is based on
minimization of the following $\chi^2$ by changing the fitting parameters (x$_i$) with their
increments $\Delta$x$_i$,

\begin{equation}\label{chisquare}
\chi^2=\sum_{i=1}^{N}\left(\frac{y_i-y([Fe/H],T_{\rm eff},\log g,V_{rot} sin(i),\zeta)}{\sigma_i} \right)^2,
\end{equation}
where $y_i$ denotes the observed quantities, $y(x_i)$ are the calculated models as a function of
atmosphere parameters and $\sigma_i$ is the standard deviation of the observed spectrum
with $N$ data points. Since the wavelength range of spectral orders we study are relatively small
($\langle$100 \AA), we assume that the standard deviation along the spectral order remains
the same ($\sigma^2_i$=$\sigma^2$) and is related with the S/N ratio of the spectrum with
$\sigma^2=\frac{1}{(S/N)^2}$. Uncertainties of the model parameters are estimated from the standard
deviation of the parameters, which are obtained for each spectral order, from the mean.

A grid of atmosphere models has been constructed for the following ranges of $[Fe/H] =
[-0.04, 0.20] (0.04) dex$, $T_{\rm eff1}$ = [16000, 19000] (200) K, $T_{\rm eff2}$ = [13000, 15000]
(200) K, $\log g_{\rm 12}$ = [4.0, 4.3] (0.1) cgs, $\zeta_{\rm 12}$ = [2, 4] (1) km\,s$^{-1}$,
V$_{\rm rot1}$~sin$i$= [110, 150] (5) km\,s$^{-1}$ and V$_{\rm rot2}$~sin$i$= [60, 100] (5)
km\,s$^{-1}$ with the increments given in brackets. In order to shrink the parameter space
in Eq.~\ref{chisquare}, the metallicity and microturbulance velocity intervals are
adopted from the spectroscopic study of four B-type Ori~OB1a members \citep{cunha} and
statistical study of B-type stars \citep{nieva2010,kodaira} respectively. The effective
temperature intervals are determined using the colour information of the system (see \S 4.1).
Surface gravity and projected rotational velocity intervals are adopted from the
components' spectral types which are estimated from their temperatures and mass functions
obtained from the spectroscopic orbit by \citet{bakis2010}.

The normalized flux of the LC of IM~Mon at out of eclipses is changing with phase due to the
elongated shapes of the component stars and due to the reflection effect, which causes a variation
in the contribution factor of the components in the total light, therefore in the composite
spectrum. In the eclipse phases, a light dilution of the eclipsed component
should also be taken into account as an additional effect on the light contributions. Therefore,
the most suitable orbital phases to study the spectral lines of the components in a binary system
with circular orbit are the quadrature phases, where components' spectral lines are the strongest
and well-seperated, unless there is a total eclipse in the system. In case of a total
eclipse, the most suitable phase to study the eclipsing component is the middle of the eclipse
phase, where the totally eclipsed component has no light contribution to the total light. In our
case, before fitting the models to the observed spectrum at a chosen orbital phase
($\phi$=0.75), the synthetic spectrum of the components is re-scaled by
considering their light contribution in each photometric band listed in table \ref{lcsolutions}.
For the spectroscopic regions different from the effective wavelengths of the photometric bands of
the observations (\textit{BVH$_p$}), an interpolation of the light contributions for a specific
wavelength of the spectral line being analyzed is applied. The Doppler shifts of the
components due to the orbital motion are also calculated using the orbital parameters given in
table 4 of \citet{bakis2010} and this shift is applied to the synthetic spectrum of the component
stars before forming the composite spectrum.

The continuum normalization procedure is also an important factor since a wrong normalization may
alter the line depths leading to an additional uncertainty in the modelling parameters. Toward the
early-type stars, the continuum is more visible due to a decreasing number and strength of
neutral metallic lines. However, the fast rotation of early-type stars also has a negative
effect on the visibility of the continuum. Therefore, those studying early-type stellar
spectrum are luckier than those studying late-type stars in the sense of determination of
continuum regions of the spectrum. The general procedure of continuum normalization of
IM~Mon spectra is already given by \citet{bakis2010}. To estimate uncertainty raising from
continuum normalization, each spectral order is nomalized five times and the
resulting line depths of the spectral lines are investigated. The standard deviation of
measured line depths is of the order of 0.1 per cent which is in the uncertainty box of
an observed spectrum with average S/N ratio of 100. Therefore, uncertainties in
our model parameters are mostly due to the bias of the spectrum rather than the continuum
normalization procedure.

\begin{figure*}
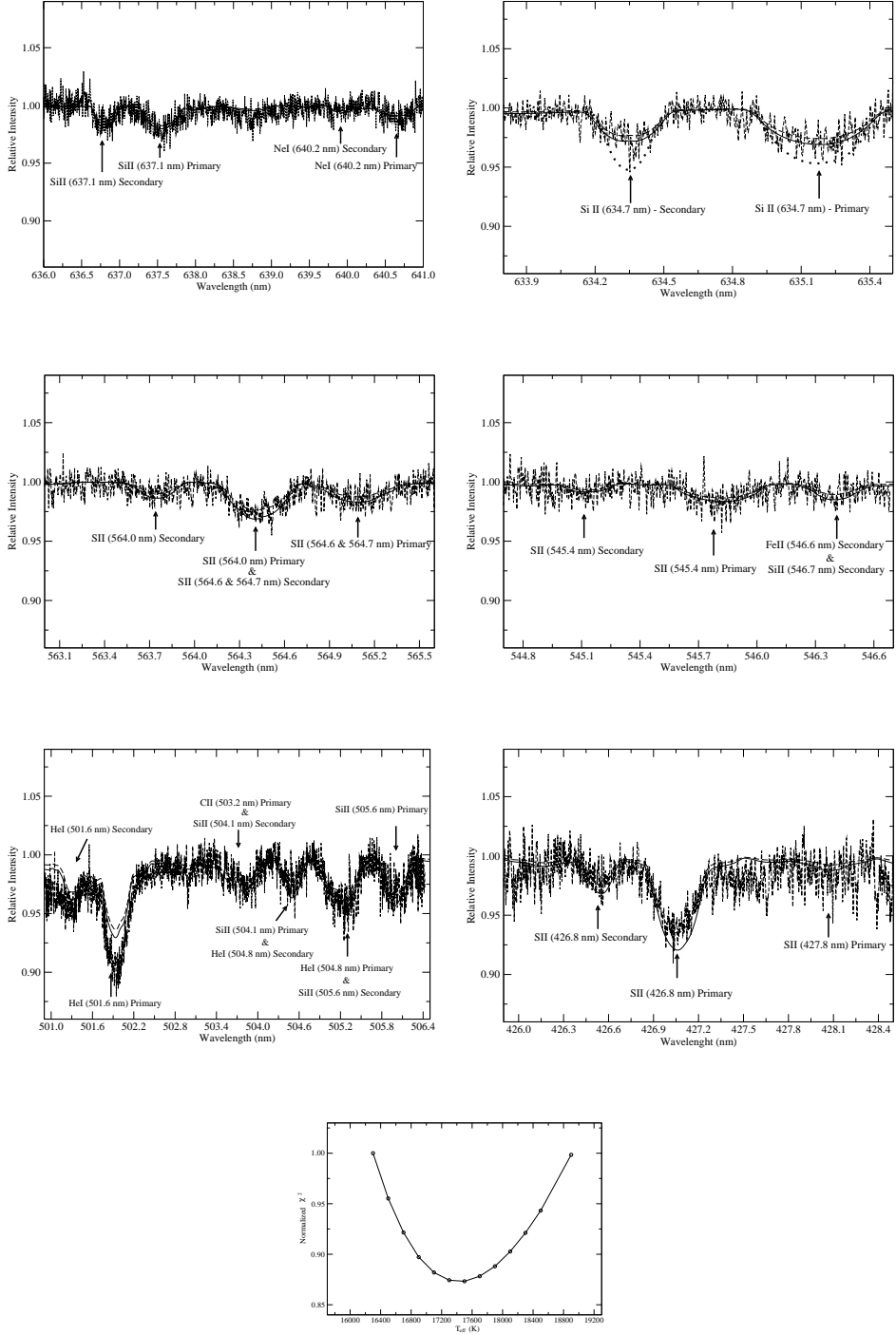

\begin{center}
\begin{tabular}{cc}
\FigureFile(60mm,60mm){figure1a.eps} & \FigureFile(60mm,60mm){figure1b.eps} \\\\
\FigureFile(60mm,60mm){figure1c.eps} & \FigureFile(60mm,60mm){figure1d.eps} \\\\
\FigureFile(60mm,60mm){figure1e.eps} & \FigureFile(60mm,60mm){figure1f.eps} \\\\
\multicolumn{2}{c}{\FigureFile(43mm,60mm){figure1g.eps}}
\end{tabular}
\caption{Observed composite spectrum of IM~Mon at orbital phase $\phi$=0.75 and the model spectrum Doppler shifted and re-scaled for the light contributions of the component stars. The synthetic spectrum calculated for solar metal abundance and for atmosphere parameters varied by 1-$\sigma$ are also shown for comparison. In each panel the dashed-dotted lines are the observed spectrum whereas the solid, dashed and dotted lines (only in top right panel) stand for the best fitting, solar metal abundance and 1-$\sigma$ varied synthetic spectrum (only for Si II 634.7 nm line), respectively. The normalized $\chi^2$ variation of primary star temperature in our grid search is shown at the bottom.} \label{atmfit}
\end{center}
\end{figure*}

All spectral lines which are clearly visible in the observed spectrum and are given in
table \ref{speclines} are analyzed. Atmospheric parameters obtained from metallic lines are
consistent for both components while for Balmer and helium lines of the primary component
a discrepancy between model and observations is noticed. This is because of the fact that starting
from late B type stars (T$_{\rm eff}$ $>$ 15000~K) there is a discrepancy between the non-LTE and
LTE models for neutral helium and Balmer lines, which progressively becomes more important as the
effective temperature of the star increases \citep{omara}. Nevertheless, the discrepancy between
LTE and non-LTE for metallic lines is negligible \citep{nieva2007}. The inconsistency of
atmospheric parameters from the metallic lines and the helium lines is due to the departure from
LTE for the primary component which is negligible or undetectable for the secondary component
because of its lower temperature. Since our atmosphere calculations are all LTE-based, we conclude
that the modelling of Balmer and helium lines of the primary component can not be performed
with our adopted LTE atmosphere models. However, the existance of the metallic lines of the
primary in the spectrum is still adequate for the determination of its atmosphere
parameters. Therefore, we used only metallic lines given in table \ref{speclines} for finding the
best fitting atmosphere parameters. The best fitting model spectrum is
shown in figure\,\ref{atmfit}.

The modelling of the observed spectrum yielded the following atmosphere parameters with their uncertainties in brackets for the components of IM~Mon; T$_{\rm eff1}$ = 17500(350)~K, T$_{\rm eff2}$ = 14500(450)~K, $\log g_{\rm 1}$ = 4.20(0.10) cgs, $\log g_{\rm 2}$ = 4.20(0.10) cgs, $\zeta_{\rm 12}$ = 2(2) km\,s$^{-1}$, $V_{\rm{rot}1}\sin i = 130(10)$ km\,s$^{-1}$, $V_{\rm{rot}2}\sin i = 80(20)$ km\,s$^{-1}$ and $[Fe/H] = 0.20(0.15)$ dex.

For comparison with the best fitting atmosphere model, we computed two more synthetic spectra: one with the solar metallicity and one with the atmosphere parameters varied by 1-$\sigma$, both are shown in figure\,\ref{atmfit}. The atmosphere parameters varied by 1-$\sigma$ are T$_{\rm eff1}$ = 17150~K, T$_{\rm eff2}$ = 14050~K, $\log g_{\rm 1}$ = 4.10 cgs, $\log g_{\rm 2}$ = 4.10 cgs, $\zeta_{\rm 12}$ = 0 km\,s$^{-1}$, $V_{\rm{rot}1}\sin i = 120$ km\,s$^{-1}$, $V_{\rm{rot}2}\sin i = 60$ km\,s$^{-1}$ and $[Fe/H] = 0.35$ dex.

The derived metallicity ($[Fe/H] = 0.20(0.15)$ dex) is just at the edge of the grid range, but the value is well-determined because the test calculation for comparison with 1-$\sigma$ variation (see figure\,\ref{atmfit}) shows a large difference between observed and theoretical spectra.

\begin{table}
\begin{center}
\caption{Spectral lines visible in the observed spectrum of IM~Mon.}\label{speclines}
\begin{tabular}{clc}
\hline\hline
 \quad\quad Spectral Order & Spectral Line & Wavelength (nm) \\
  \hline
85       & He I           & 667.8     \\
87       & H$_\alpha$     & 656.3     \\
89       & Ne I           & 640.2     \\
         & Si II          & 637.1     \\
90       & Si II          & 634.7     \\
97       & He I           & 587.5     \\
101      & S II           & 564.7     \\
         &                & 564.6     \\
         &                & 564.0     \\
104      & Si II          & 546.7     \\
         & Fe II          & 546.6     \\
         & S II           & 545.4     \\
113      & He I           & 504.8     \\
         & Si II          & 504.1     \\
         & C II           & 503.2     \\
         & He I           & 501.6     \\
116      & S II           & 492.5     \\
         & Fe II          & 492.4     \\
         & He I           & 492.2     \\
121      & He I           & 471.3     \\
127      & Mg II          & 448.1     \\
         & He I           & 447.2     \\
130      & He I           & 438.8     \\
133      & S II           & 426.8     \\
137      & He I           & 414.4     \\
141      & He I           & 402.6     \\
\hline
\end{tabular}
\end{center}
\end{table}

\section{Close Binary Stellar Parameters}

\subsection{Binary Model and Input Parameters}

Accurate estimation of the primary star temperature is one of the most critical tasks before the LC
modelling of close binary stars. A wrong estimation affects the determination of the secondary
star's temperature which leads to incorrect evolutionary scenarios. In this study two
methods are used for temperature estimation, a) by modelling spectral lines and b) by the
$Q$-method of \citet{johnson1953} using the $UBV$ colours of IM~Mon. The modelling of spectral
lines is already discussed in \S 3 and the primary star temperature is found to be T$_{\rm
eff1}=17500$ K. The $Q$-method of \citet{johnson1953} is based on linear correlation between the
reddened and unreddened coulor of stars as the following relation:

\begin{equation}\label{qmethod}
Q=(U-B)-\frac{E(U-B)}{E(B-V)}(B-V),
\end{equation}
where $(U-B)$ and $(B-V)$ are the reddened colour indices and $E(U-B)$ and $E(B-V)$ are colour
excesses in these indices respectively. \citet{johnson1953} determined the mean value of
$\frac{E(U-B)}{E(B-V)}$ to be 0.72$\pm$0.03 from the unreddened and reddened stars tabulated in
their study. Since the linear correlation between the reddened and unreddened colours is valid for
early-type stars, the usage of the $Q$-method is limited for the spectral type range B1-B9 which
corresponds to range in the $Q$ parameter as --0.80$<Q<$--0.05. Once the $Q$-parameter is obtained,
the unreddened colour index $(B-V)_0$ can be derived from $(B-V)_0=-0.009+0.337Q$. Using
the unreddened colour index of $(B-V)_0$, one can derive the colour excess of $E(B-V)$
from $E(B-V)$=$(B-V)_0-(B-V)$. The unreddened colour excess $E(U-B)$ can also be derived from the
relation $\frac{E(U-B)}{E(B-V)}=0.72\pm0.03$. Then the unreddened colour index of
$(U-B)_0$ is found from $E(U-B)$=$(U-B)_0-(U-B)$.

The combined colours of IM~Mon were collected from \citet{deutschman} as $(U-B)=-0.650(0.013)$ and
$(B-V)=-0.150(0.019)$. The combined colour of IM~Mon yielded the $Q$-parameter $Q=-0.542\pm0.03$,
colour excess $E(B-V)=0.042$$\pm$0.033 and visual absorption $A_{\rm v}=0.129$. Therefore, the
unreddened colours of the system are $(B-V)_0 =$ --0.192$\pm$0.014 mag and
$(U-B)_0=-0.680$$\pm$0.039 mag. These unreddened colours correspond to a temperature of
17000$\pm$200~K according to the calibration tables of \citet{cramer}. However, it should be noted
that these colours and the corresponding temperature are obtained from the combined light of the
components, resulting in a slightly redder colour and lower temperature than the intrinsic
colour of the primary component, although the LCs and the spectral lines show that the light of the
primary star dominates. Since the spectral line modelling yields intrinsic temperatures of the
component stars, we adopted a temperature of T$_{eff1}$=17500~K$\pm$350~K for the primary star.

\subsection{Determination of the Photometric Elements}

From table \ref{Listsources}, where all available photometric data of IM~Mon are listed,
five photometric data sets (\textit{B} band data of \citet{gum}, \textit{B} and \textit{V} band
data of Sanyal et al. (1965), \textit{V} band data of \citet{shobbrook} and H$_p$ band data of
Hipparcos) are selected for simultaneous analysis of LCs together with the RVs of the components
using the 2003 version of the Wilson-Devinney LC analysis code (Wilson \& Devinney, 1971; Wilson,
1994). Selection was made according to their relative precision among others.

Using the best data specified above we obtained the estimate of the period $P$ and initial
epoch of primary minimum $M_0$ together with the model LCs in various colours and model RV curve.
We then found that there are systematic seasonal differences between the observed and the model
LCs. The seasonal LC variations were already mentioned by \citet{sanyal1964} who observed
the system extensively during 1962--1964. They found that the scatter seen in the light curve of
IM~Mon is not random but systematic due to intrinsic variability of the cooler component. The
authors argued that the period of these variations is nearly, but not exactly, equal to
half of the orbital period. Such type of variability should then manifests itself in
long-term changes in the shape of LC in individual observational sets.

\begin{figure}
\centering \resizebox{0.7\hsize}{!}{\includegraphics{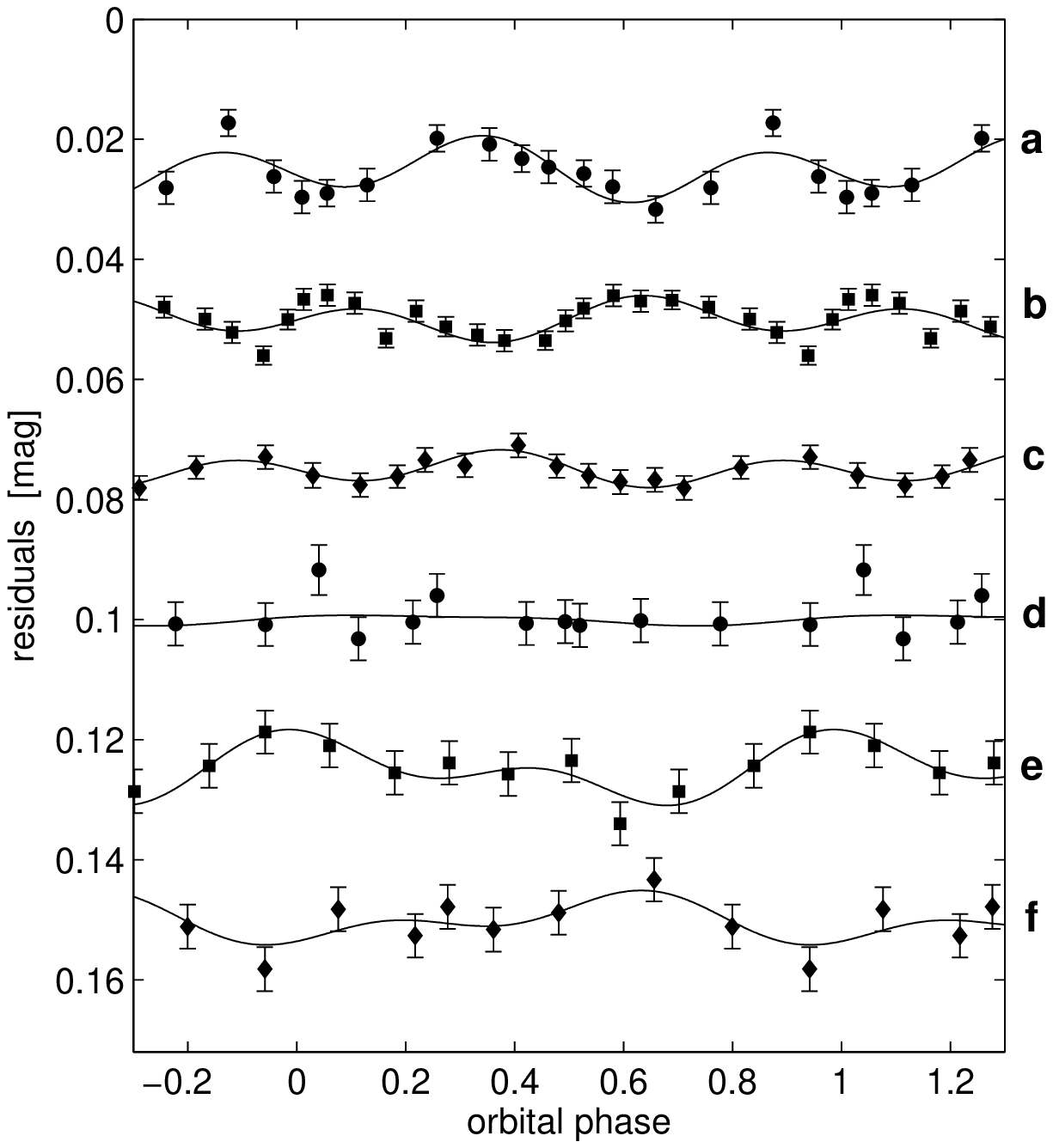}} \caption{Seasonal variability of
light curve shapes visualized through the difference of an individual observed light curve
defined by their normal points and mean light curve. a -- \citet{gum}, b -- \citet{sanyal1965}, c
-- \citet{esa}, d -- \citet{shobbrook}, e -- \citet{pojm97}, f -- \citet{malek}. }\label{inter}
\end{figure}

The presence of LC changes is clearly apparent in Fig.\ref{inter} depicting the difference
between observed LCs and the mean LC. The double-wave character of these differential LCs
supports the findings of \citet{sanyal1964}, however their hypothesis on the
intrinsic variability of the cooler component is not unique. Observed seasonal LCs changes
resemble those ones found in the non-eclipsing interacting binary HD~143\,418 containing a
subsynchronously rotating primary passing through its synchronization stage. The seasonal
variability of the orbitally modulated light curves is related to an expected incidence of
circumstellar matter originating in the tidally spinning up primary component \citep{bozic,zverko}.
This makes the more detailed study of seasonal changes of IM Mon appealing.

Aiming to restrain the influence of seasonal variations we corrected all observed data by
approximating the data by the double wave harmonic function. The scatter of residuals then
diminished considerably. Subsequently, the new model phase curves were used as templates
that were applied to all the observational data summarized in the table\,\ref{Listsources} with
the aim to compute improved ephemeris.

The procedure is iterative, assuming that $y_i$ is $i$-th measurement obtained in the time $t_i$,
$F(\vartheta(t_i))$ is the model prediction for the $i$-th measurement and $w_i$ is its weight
(inversely proportional to the square of its assumed uncertainty). $\vartheta(t_i)$ is then
the so-called phase function (the fractional part of it being the common phase $\varphi$,
the integer part is the epoch $E$). The linear phase function is determined by the simple
relation: $\vartheta(M_0,\,P,\,t)=(t-M_0)/P$, where $M_0$ is the time of initial epoch of
primary minimum and $P$ is the orbital period. We found parameters $P$ and $M_0$ so that
the following is valid:

\begin{equation} \label{ephem}
\Delta y_i= y_i - F(\vartheta(t_i)),\quad \sum_{i=1}^n \Delta y_i
\frac{\partial F}{\partial \vartheta_i}\, w_i=0\ \textrm{and}\
\sum_{i=1}^n \Delta y_i \frac{\partial F}{\partial \vartheta_i}
\vartheta_i\,w_i=0.
\end{equation}
After several iterations we obtained the following equation for heliocentric data of the primary minima:
\begin{equation} \label{linephem}
\mathrm{HJD_I}=M_0+P\,E=2\,442\,331.2515(9)+1\timeform{d}.19024249(14)\,E,
\end{equation}
where $E$ is an epoch number.

The orbital period $P=1\timeform{d}.19024249(14)$ given in eq.\,\ref{linephem} agrees with the
period $P=1\timeform{d}.1902424$ found by \citet{kreiner}. Nevertheless, the present value of the
orbital period is more reliable and precise because it is based on much larger observational
material and on their more careful treatment.

Supposing a non-zero but constant time derivative of the period $\dot{P}$ we found a
moderate increase of the orbital period of $\dot{P}=(9\pm5)\times 10^{-11}$, which may not be real.

The orbital period ($P$) and the initial epoch of the primary minimum ($M_0$) were kept
fixed during the simultaneous solutions. Since the time derivative of the orbital period
($\dot{P}$) is statistically not significant, its value is taken to be zero as the fixed
parameter. The temperature of the primary was fixed at $T_{\rm eff1}$ = 17500~K and the
temperature of the secondary ($T_{\rm eff2}$) was left to converge. Gravity darkening exponents
$g_{\rm 1}=g_{\rm 2}=1$ and bolometric albedos $A_{\rm 1}=A_{\rm 2}=1$ were set for radiative
envelopes \citep{vonzeipel}. The logarithmic limb-darkening law was used and
limb-darkening coefficients were taken from \citet{vanhamme}. The surface potentials ($\Omega_{\rm
1,2}$), light factors of the components ($l_{\rm 1,2}$), orbital inclination ($i$), the mass ratio
($q$), seperation ($a$) of the components and the systemic velocity ($V_\gamma$) were the adjusted
parameters during the modelling of the LCs. The orbital eccentricity ($e$) and longitude of
periastron ($\omega$) were fixed for a circular orbit \citep{bakis2010}.

In addition to the LC input parameters given above, the solution mode, which describes the type of
the binary (i.e. detached binary, semi-detached binary, contact binary), is required by the
Wilson-Devinney code as program input. The most recent photometric analysis of IM~Mon by
\citet{cester} was not succesfull to classify the type of the binary due to large scatter in the
photometric data they used. Nevertheless, \citet{pourbaix} emphasized the possibility of contact
status of the system in \textquotedblleft SB9: The ninth catalogue of spectroscopic binary orbits \textquotedblright. We, therefore,
initially tested all possible solution modes (Mode 2 for detached binaries, Mode 3 for overcontact
binaries with components having different surface brightnesses, Mode 4 and 5 for semi-detached
binaries and Mode 6 for double contact binaries) with the Wilson-Devinney code. None of the
solutions with contact and semi-contact configurations converged to give a good fit and in each
trial the fitting parameters converged to a detached configuration. Hence, in the following steps
of the analysis, detached binary configuration has been adopted.

\begin{figure}
\begin{center}
\vspace{1cm}
\begin{tabular}{c}
\FigureFile(80mm,80mm){figure3a.eps} \\\\
\FigureFile(90mm,90mm){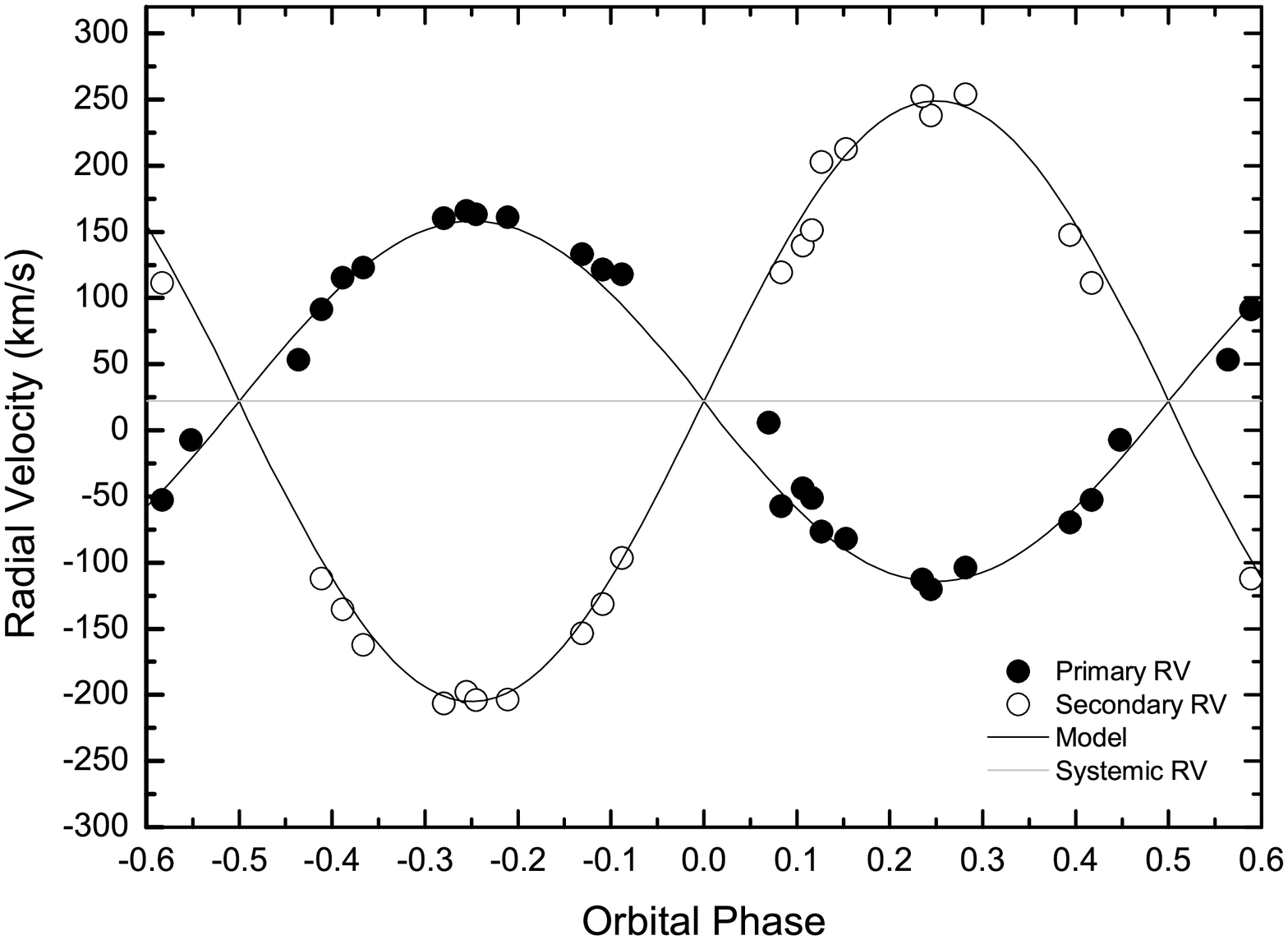} \\\\
\FigureFile(80mm,80mm){figure3c.eps} \\
\end{tabular}
\caption{Theoretical model fits to five different photometric band LCs (\textit{top}) and RV data (\textit{middle}) of IM~Mon. The star shapes are plotted for different orbital phases (\textit{bottom}).} \label{BVRfit}
\end{center}
\end{figure}

The solutions with detached configuration in each photometric band converged very rapidly, and had
the smallest residuals. Input values of the adjusted parameters and the primary star temperature
were then altered to check the consistency and uniqueness of the solution. A change in the value of
primary star temperature did not affect other parameters' output values significantly
except for the secondary star temperature. These new input parameters converged to the
parameters of the first solution, which shows the consistency of the solutions listed in table
\ref{lcsolutions}. The uncertainties of the final light curve modelling parameters (see
table~\ref{lcsolutions}) directly come from the light curve modelling program output. Adopted LC
solutions for each photometric set are shown in figure\,\ref{BVRfit} together with RV
solutions and the star shapes at four different orbital phases. The Rossiter-McLaughlin effect seen
near the eclipse phases of close binary systems is very small in the case of
IM~Mon due to the low inclination of its orbit (see star shapes at eclipses in
figure\,\ref{BVRfit}). However, near the primary eclipse, there is one of the RVs of the
primary component significantly outlying from the theoretical curve. This RV is clearly not
measured precisely due to the blending of the spectral lines at this phase and the scatter is
certainly not due to Rossiter-McLaughlin effect since it is red-shifted.

\begin{table*}
\begin{center}
\small \caption{Results from the simultaneous solution of {\em B}, {\em V} and {\em Hp}-band LCs of IM~Mon system. Adjusted and fixed parameters are presented in separate panels of the table. Uncertainties of adjusted parameters are given in brackets.} \label{lcsolutions}
\begin{tabular}{lc}\hline\hline
Parameter              &  Value         \\
\hline
\multicolumn{2}{l}{Adjusted parameters:}\\
\hline
$T_{\rm eff2}(K)$      &   14500(200)   \\
$L_{1}/L_{1+2}(B)$     &   0.700(0.003) \\
$L_{1}/L_{1+2}(Hp)$    &   0.698(0.005) \\
$L_{1}/L_{1+2}(V)$     &   0.693(0.002) \\
$\Omega_{\rm 1}$       &  3.75(0.12)    \\
$\Omega_{\rm 2}$       &  3.72(0.08)    \\
$\Omega_{\rm cr}$      &  3.07          \\
$r_{\rm 1}$(mean)      &   0.323(0.020) \\
$r_{\rm 2}$(mean)      &   0.242(0.030) \\
$i (^{o})$             &   62.2(0.9)    \\
$q$                    &   0.603 (0.011)\\
$a (R_\odot)$          &   9.77 (0.14)  \\
$V_{\gamma}$ (km\,s$^{-1}$) & 22.1(2.1) \\
\hline
\multicolumn{2}{l}{Fixed parameters:}   \\
\hline
$P$                    & 1$\timeform{d}$.19024249  \\
$\dot{P}$              & 0.0            \\
$M_0$                  & 2\,442\,331.2515  \\
$e$                    & 0.0            \\
$A_1=A_2$              & 1.0            \\
$g_1=g_2$              & 1.0            \\
$T_{\rm eff1}(K)$      &   17500        \\
x$_{\rm 1}(B,V,H_p)$     & 0.516, 0.438, 0.462  \\
y$_{\rm 1}(B,V,H_p)$     & 0.279, 0.234, 0.249  \\
x$_{\rm 2}(B,V,H_p)$     & 0.561, 0.478, 0.504  \\
y$_{\rm 2}(B,V,H_p)$     & 0.303, 0.254, 0.269  \\
$F_{\rm 1}$=$F_{\rm 2}$     & \multicolumn{1}{c}{1.0} \\
\hline
$\chi^2_{\rm min}(B_1,B_2, V_1, V_2, H_p)$   &    0.0029, 0.0032, 0.0039, 0.0056, 0.0042 \\
\hline
\multicolumn{2}{l}{$B_1$ (Gum, 1951), $B_2$ (Sanyal et al., 1965),}  \\
\multicolumn{2}{l}{$V_1$ (Sanyal et al., 1965), $V_2$ (Shobbrook, 2004), $H_p$ (ESA, 1997).}
\end{tabular}
\end{center}
\end{table*}

\section{Results and Discussion}

\subsection{Astrophysical Parameters}

The fundamental astrophysical parameters of IM~Mon, which were derived from the simultaneous solutions of light and RV curves (see \S4.2), are summarized in table~\ref{astparams}. The parameter uncertainties in table~\ref{astparams} were computed by means of applying the laws of error propagation based on the formal errors given in table~\ref{lcsolutions}. The temperature $T_{\rm eff1} =$ 17500~K, mass $M_{\rm 1}=5.50 M_{\odot}$ and radius $R_{\rm 1}=3.15R_{\odot}$ of the primary component correspond to a spectral type of B4V. The temperature $T_{\rm eff2} =$ 14500~K of the secondary star implies a B6 spectral type ZAMS star while its mass $M_{\rm 2}=3.32 M_{\odot}$ and radius of $R_{\rm 2}=2.36R_{\odot}$ are more consistent with a spectral type of B6.5 ZAMS star \citep[i.e.][]{straizys}.

The unreddened Johnson \textit{V}-magnitude \citep{deutschman} of IM~Mon, when combined with the
light contributions (see table~\ref{lcsolutions}) as derived from the light-curve analysis, yields
the intrinsic \textit{V}-magnitudes $m_{\rm V1}$=6.84 mag and $m_{\rm V2}=$7.72 mag of the primary
and secondary components, respectively. Using bolometric corrections $BC_{\rm 1}=$--1.60(0.04) and
$BC_{\rm 2}=$--1.16(0.08) mag for the primary and the secondary \citep[i.e.][]{flower1996},
bolometric and absolute visual magnitudes of the close binary components are derived (see table
\ref{astparams}). In \citet{flower1996}, bolometric corrections are not listed with their
uncertainties. We estimated the uncertainties by using the uncertainties of the temperatures of the
component stars. Considering the uncertainties coming from the bolometric corrections and visual
magnitude of the system, the distance modulus indicates a photometric distance of 353(59) pc to
IM~Mon, which is larger than the distance (304(36) pc) derived for IM~Mon by \citet{vanleeuwen}
from the re-analysis of raw Hipparcos data, but not inconsistent with it in view of the large error
bar of both distance determinations. In the original Hipparcos catalogue, the distance to IM~Mon is
given as 341(85) pc.

The projected rotational velocities of the components were derived to be $V_{\rm rot1}$~$\sin(i)$ = 130(10) km\,s$^{-1}$ and $V_{\rm rot2}$~$\sin(i)$ = 80(10) km\,s$^{-1}$ in \S~3. Using the orbital inclination ($i=62^{\circ}.2$) of IM~Mon, the observed rotational velocities of the components are found to be $V_{\rm rot1}$ = 147 (15) km\,s$^{-1}$ and $V_{\rm rot2}$ = 90 (25) km\,s$^{-1}$. Both components' rotational velocities seem to agree with the theoretical synchronization velocities within their error limits.

\begin{table*}
\begin{center}
\small \caption{Close binary stellar parameters of IM~Mon. Errors of parameters are given in parenthesis.} \label{astparams}
\begin{tabular}{lccc}\hline
Parameter                          & Symbol  & Primary & Secondary                    \\
\hline
Spectral type                      & Sp      & B4 V & B6.5 V                       \\
Mass (M$_\odot$)                   & \emph{M}       & 5.50(0.24) & 3.32(0.16)         \\
Radius (R$_\odot$)                 & \emph{R}       & 3.15(0.04) & 2.36(0.03)         \\
Separation (R$_\odot$)             & \emph{a}       & \multicolumn{2}{c}{9.77(0.14)}   \\
Orbital period (days)              & \emph{P}       & \multicolumn{2}{c}{1.19024249(14)}   \\
Orbital inclination ($^{\circ}$)   & \emph{i}       & \multicolumn{2}{c}{62.2(0.9)}   \\
Mass ratio                         & \emph{q}       & \multicolumn{2}{c}{0.603(0.011)}\\
Eccentricity                       & \emph{e}       & \multicolumn{2}{c}{0.0}\\
Surface gravity (cgs)              & $\log g$       & 4.181(0.009)& 4.214(0.015)      \\
Integrated visual magnitude (mag)  & \emph{V}       &  \multicolumn{2}{c}{6.57(0.03)} \\
Individual visual magnitudes (mag) & \emph{V}       &  6.84(0.03) & 7.72(0.03)\\
Integrated colour index (mag)      & $B-V$         &  \multicolumn{2}{c}{-0.15(0.02)} \\
Colour excess (mag)                &$E(B-V)$&\multicolumn{2}{c}{0.04(0.03)}\\
Visual absorption (mag)            & $A_{\rm v}$ &\multicolumn{2}{c}{0.13(0.03)}\\
Intrinsic colour index (mag)       & $(B-V)_{\rm 0}$&\multicolumn{2}{c}{-0.19(0.02)}\\
Temperature (K)                    & $T_{\rm eff}$ & 17500(350) & 14500(550) \\
Luminosity (L$_\odot$)             & $\log$ \emph{L}& 2.92(0.03) & 2.34(0.06)\\
Bolometric magnitude (mag)         &$M_{\rm bol}$& -2.55(0.08) & -1.11(0.16)    \\
Absolute visual magnitude (mag)    &$M_{\rm v}$  & -0.97(0.04) & -0.02(0.07)    \\
Bolometric correction (mag)        &\emph{BC}& -1.60(0.04)      & -1.16(0.08)   \\
Velocity amplitudes (km\,s$^{-1}$)  &$K_{\rm 1,2}$& 138.7(3.1) & 228.8(3.1)    \\
Systemic velocity (km\,s$^{-1}$)    &$V_{\gamma}$ & \multicolumn{2}{c}{22.1(2.1)} \\
Computed synchronization velocities (km\,s$^{-1}$)& V$_{synch}$ & 134(2) & 100(2) \\
Observed rotational velocities (km\,s$^{-1}$) & V$_{\rm rot}$ & 147(15) & 90(25) \\
Distance (pc)                      &\emph{d} & \multicolumn{2}{c}{353(59)} \\
Proper motion (mas yr$^{-1}$) &$\mu_\alpha cos\delta$, $\mu_\delta$ &
\multicolumn{2}{c}{0.00(0.41), 3.66(0.33)$^{*}$} \\
Space velocities (km\,s$^{-1}$) & $U, V, W$ &
\multicolumn{2}{c}{-21.80(1.79), -7.17(1.22), -0.12(0.72)}\\
\hline
* from {\em Hipparcos} catalogue \citep{vanleeuwen}.
\end{tabular}
\end{center}
\end{table*}

\subsection{Kinematical and Dynamical Analysis}

The kinematical properties of IM~Mon have been derived by means of studying IM~Mon's space velocity
which was calculated using the algorithm given by \citet{johnson1987}. Calculation of space
velocities requires knowing the systemic velocity and distance of IM~Mon as well as its
proper motion. The systemic velocity and distance of IM~Mon have been derived in
the present study and are presented in table~\ref{astparams}. The proper motion data were taken
from the newly reduced Hipparcos catalogue of \citet{vanleeuwen}. The $U$, \textit{V} and
$W$ space velocity components and their errors are listed in table \ref{astparams}. To obtain the
space velocity precisely, the first-order galactic differential rotation correction (DRC) was taken
into account \citep{mihalas}, and -5.19 and -0.72 km\,s$^{-1}$ DRCs were applied to $U$ and
\textit{V} space velocity components respectively. The W velocity is not affected in this
first-order approximation. As for the local standard of rest correction, \citet{mihalas} values (9,
12, 7) km\,s$^{-1}$ were used and the total space velocity of the system was obtained as
$S=22.3$(2.3) km\,s$^{-1}$. To determine the population type of IM~Mon, the galactic orbit of the
system was examined. Using the N-body code of \citet{dinescu}, the system's apogalactic ($R_{\rm
max}$) and perigalactic ($R_{\rm min}$) distances were obtained as 8.64 and 8.25 kpc, respectively.
Also, the maximum possible vertical distance of the system from the galactic plane is $|z_{\rm
max}|=|z_{\rm min}|=90$ pc. The following formulae were used to derive the planar and vertical
ellipticities:

\begin{equation}
e_{p}=\frac{R_{max}-R_{min}}{R_{max} + R_{min}},
\end{equation}

\begin{equation}
e_{v}=\frac{|z_{max}|+|z_{min}|}{R_{max}+R_{min}}.
\end{equation}
The planar and vertical ellipticities were calculated as $e_{p}=0.02$ and $e_{v}=0.01$. These values show that IM~Mon is orbiting around the center of the Galaxy in a circular orbit and the system belongs to the young thin-disc population.

\subsection{Evolutionary Stage}

\begin{figure}
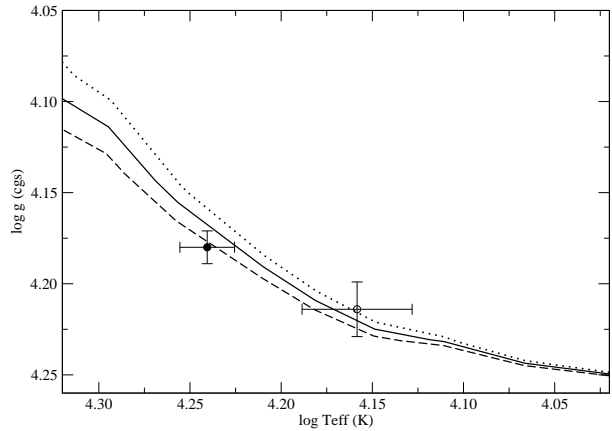

\begin{center}
\vspace{1cm}
\begin{tabular}{c}
\FigureFile(80mm,80mm){figure4.eps} \\
\end{tabular}
\caption{Isochrones for IM~Mon calculated metallicity for $[Fe/H]=0.2$ dex. The dotted, solid and dashed lines stand for 12, 11.5 and 11 Myr, respectively. The primary and secondary components are shown with filled and empty circles, respectively.} \label{isochrones}
\end{center}
\end{figure}

We investigated the evolutionary status of IM~Mon in the plane of $\log T_{\rm eff}$ -
$\log g$ (figure\,\ref{isochrones}) using the latest theoretical isochrones of \citet{girardi},
which include mass loss and moderate overshooting ($\Lambda_c=$0.5). Assuming a $[Fe/H]=$0.2 dex
metal content as obtained from the modelling of spectral lines, we prepared a set of isochrones
corresponding to $Y=0.30$ and $Z=0.03$. The isochrones of 11 Myr and 12 Myr shown in
figure\,\ref{isochrones} imply a mean age of 11.5(1.5) Myr for the system. The location of both
components of IM~Mon is fully compatible with the formerly derived ages for Ori~OB1a which is
summarized in table \ref{oriob1} together with other related parameters.

\subsection{Membership to Ori~OB1a}

To study the membership of IM~Mon to the Ori~OB1a association the galactic orbits of known
associated stars were generated. \citet{brown1994} listed the stars that are secure
members of Ori~OB1a association. To generate more precise galactic orbits, the proper motions and
trigonometric parallaxes of member stars were taken from the newly reduced Hipparcos catalogue
\citep{vanleeuwen}. 29 members of Ori~OB1a with both astrometric data and precise RVs
\citep{kharchenko} were found.

\begin{figure}
\begin{center}
\vspace{1cm}
\begin{tabular}{c}
\FigureFile(80mm,80mm){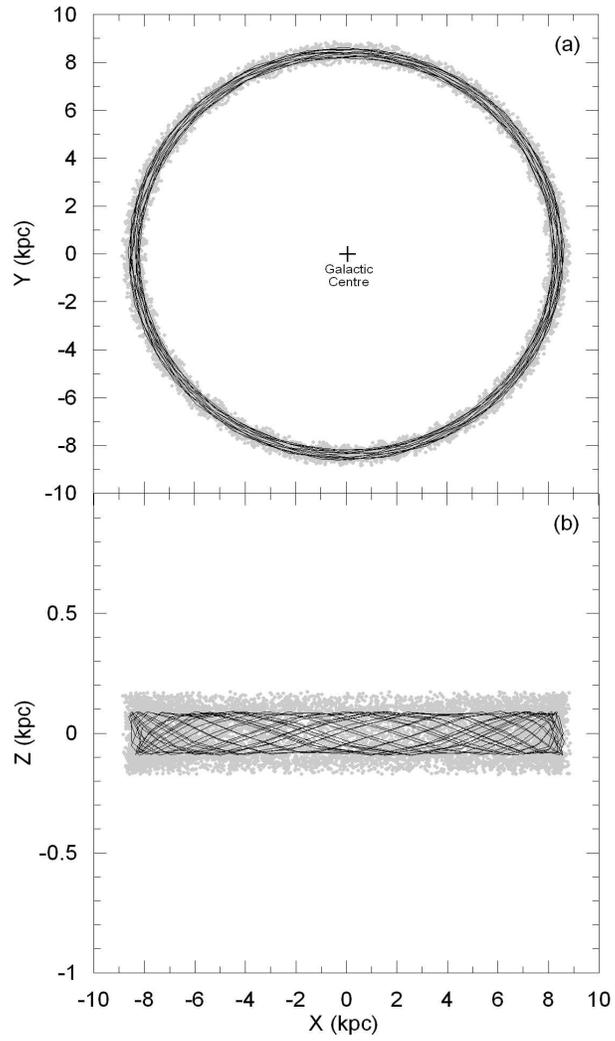} \\
\end{tabular}
\caption{The orbital motions of Ori~OB1a members (gray dots) and IM~Mon (solid line) on the projections of $X-Y$ and $X-Z$ planes around the Galactic center (for 1 Gyr) } \label{galorbit}
\end{center}
\end{figure}

The galactic orbits of the 29 stars were drawn using the N-body code of \citet{dinescu}. The
timescale in generating the orbits was assumed to be 1 Gyr and the calculation steps were 5 Myr.
The 1 Gyr timescale was assumed so that precise orbits were created, even though it is longer than
the nuclear time scale of the early type stars. The motions of those 29 stars on the $X-Y$ and
$X-Z$ planes around the Galactic center are shown in figure\,\ref{galorbit}. The galactic orbits of
the member stars are shown with gray dots, whereas IM~Mon is represented with the solid line. As
seen in figure\,\ref{galorbit}, the galactic orbits of members of Ori~OB1a are in the same region
with IM~Mon. This supports the membership of IM~Mon to the Ori~OB1a association
dynamically.

The comparison of the physical properties of the IM Mon system and Ori~OB1a association
is given in table~\ref{oriob1}. IM Mon's age was calculated as 11.5(1.5) Myr using Padova
isochrones, whereas \citet{blaauw} and \citet{brown1999}, who studied stars in Ori~OB1a region,
gave 12 and 11 Myr for the association. These values agree with the values determined for IM Mon in
this study.

The distance also needs to be studied to determine the membership of IM Mon to the Ori~OB1a
association. In this study, the distance determined for IM Mon is 353(59) pc. The distance
evaluated using photometric methods for Ori~OB1a, 336(16) pc \citep{brown1999}, and the distance
evaluated using newly reduced Hipparcos data, 400 pc \citep{melnik2009}, are in agreement with the
distance of IM Mon.

Regarding the metallicity of OB1a sub-group, we have only the spectroscopic analysis
results of \citet{cunha} on four B-type stars belonging to the region of OB1a sub-group with their
metallicity values ranging more than 1-$\sigma$ (0.13 dex) around the solar metallicity.
Nevertheles, a recent study of \citet{dorazi}, who analyzed the spectra of low-mass members, gives
a more precise value for the average metallicity ($[Fe/H]=-0.01(0.04)$ dex), a solar metallicity, of
the Orion Nebular Cluster (ONC). Therefore, the metallicity obtained for IM~Mon in this study
($[Fe/H]=+0.2(0.15)$ dex) does not agree with the previously determined metallicity values in the
region. However, considering its large uncertainty, it is still close to the solar metallicity.

In the present study, we updated the systemic velocity of IM~Mon to be
V$_\gamma$=22.1(2.1) km\,s$^{-1}$ from the value (V$_\gamma$=21.2(1.8) km\,s$^{-1}$) given by
\citet{bakis2010}. \citet{brown1999} and \citet{melnik2009} who studied the RVs of stars in the
Ori~OB1a association give the mean velocities of the member stars to be 23.0 and 25.4 km\,s$^{-1}$.
These values agree with the systemic velocity of IM Mon within its uncertainty.

\begin{table*}
\begin{center}
\setlength{\tabcolsep}{5pt}
\small \caption{Comparison of Ori~OB1a and IM~Mon.} \label{oriob1}
\begin{tabular}{|c|r|c|r|c|r|c|r|}
\hline
\hline
\multicolumn{2}{|c|}{\textbf{Age (Myr)}}                &  \multicolumn{2}{c|}{\textbf{Distance (pc)}}     & \multicolumn{2}{c|}{\textbf{Chemical Abundance (dex)}}   & \multicolumn{2}{c|}{\textbf{V$_\gamma$ (km\,s$^{-1}$})} \\
\hline
This Study   & \multicolumn{1}{c|}{Literature}&  This Study   & \multicolumn{1}{c|}{Literature}& This Study   & \multicolumn{1}{c|}{Literature}&This Study & \multicolumn{1}{c|}{Literature}\\
\hline
11.5(1.5)   &  12 (a)             & 353(59)    & 336(16) (b)   & 0.20(0.15)   & --0.01(0.04) (c) &  22.1(2.1)& 23.0 (b)      \\
            &  11 (b)             &            & 400 (d)       &              &                  &           & 25.4(1.0) (d) \\
            &  [7-10]  (f, g)     &            & 304(36)(e)    &              &                  &           &               \\
\hline
\end{tabular} \\
(a) \citet{blaauw}, (b) \citet{brown1999}, (c)\citet{dorazi}, (d)\citet{melnik2009}, (e)\citet{vanleeuwen}, (f) \citet{calvet}, (g) \citet{briceno}
\end{center}
\end{table*}

\section{Summary and Conclusions}

OB associations are young galactic clusters where star formation is ongoing or has just
ended. The study of OB associations yields useful information about the characteristic of
star formation such as formation history, binary population and initial mass function. However,
this information can be obtained only if the properties of the OB association such as
distance, age, metallicity and kinematics are very well established. Observing single stars
in an OB association does not provide information of sufficient precision unless
many of them are observed, which requires a lot of observing time. In this case, eclipsing binaries
which are the royal road to the stars can yield precise age, metallicity and kinematics of
the medium in which they are embedded, and they do not require much observing time
provided that stars with relatively short periods are selected.

In this work, using sophisticated modelling tools, we studied the close binary system IM~Mon
together with all its available photometric, spectroscopic, kinematical and dynamical
data. The membership of IM~Mon to Ori OB1a sub-group has been established securely by means of
comparing the dynamical galactic orbits of 29 Ori~OB1a members with the galactic orbit of IM~Mon.
The absolute dimensions we derived for IM~Mon in this study lead to a reliable
distance determination. The location of both components of IM~Mon in the plane of $\log T_{\rm
eff}$ - $\log g$ is fully compatible with the formerly derived ages for this association.
The metallicity ($[Fe/H]=$0.20(0.15) dex) of IM~Mon obtained in this study has a large
uncertainty which may explain the disagreement between the average metallicity
($[Fe/H]=-$0.01(0.04) dex) of the ONC and IM~Mon.

Using the information derived in the present work, we conclude that Ori~OB1a is located at a distance of 353(59) pc, has an age of 11.5(1.5) Myr and has a metallicity of 0.20(0.15) dex. In summary, IM Mon is a secure member of the Ori OB1A subgroup.

\textbf{Acknowledgements}\\
This study is fully supported by The Scientific \& Technological Research Council of Turkey
(TUBITAK) with the project code 109T449. Zden\v ek Mikul\'a\v sek is supported by the grants GAAV
IAA 301630901 and GA\v{C}R 205/08/0003. We thank S. N. de Villiers for valuable comments to
our text and the anonymous referee who improved the manuscript by his/her very useful comments.

\end{document}